\documentclass[twocolumn,showpacs,preprintnumbers,prl]{revtex4}%
\usepackage{amsfonts}
\usepackage{amssymb}
\usepackage{amsmath}
\usepackage{graphicx}
\usepackage{dcolumn}
\usepackage{bm}%
\begin{document}
\title{Phase Direct CP Violations and General Mixing Matrices}
\author{Ling-Lie Chau}
\affiliation{Physics Department, University of California, Davis}

\begin{abstract}
I formulate expressions for amplitudes suitable for quantifying both modulus
and phase direct CP violations. They result in M\"{o}bius transformation (MT)
relations, which provide encouraging information for the search of direct CP
violations in general. I apply the formulation to calculate the measurements
of phase direct CP violations and strong amplitudes in $B^{\mp}\rightarrow
K^{\mp}\pi^{\pm}\pi^{\mp}$ by the Belle collaboration. For the formulation, I
show a versatile construction procedure for $N\times N$
Cabibbo-Kobayashi-Maskawa (CKM) matrices, Pontecorvo-Maki-Nakagawa-Sakata
(PMNS) matrices, and general unitary matrices. It clarifies the $3\times3$
cases and is useful for the beyond.

\end{abstract}
\maketitle

\bigskip

\textbf{Introduction --- }CP violation studies and observations have had a
long interesting history [1-11]. CP violation in $B^{\mp}$ published in [1]
was the first of its kind, direct and without particle-antiparticle
oscillation. Further, in multiparticle decays, the total
amplitudes,\ $A_{\left(  tot\right)  }$ and $\bar{A}_{\left(  tot\right)  }$,
are coherent sums of amplitudes for various final resonances and backgrounds,
$A_{\left(  k\right)  }$ and $\bar{A}_{\left(  k\right)  }$,
\begin{equation}
A_{\left(  tot\right)  }=%
{\textstyle\sum\nolimits_{k=1}^{n}}
f_{k}A_{\left(  k\right)  }\text{ \ and \ }\bar{A}_{\left(  tot\right)  }=%
{\textstyle\sum\nolimits_{k=1}^{n}}
f_{k}\bar{A}_{\left(  k\right)  }\text{ ,} \label{Asum}%
\end{equation}
where $f_{k}$ are functions of invariant masses of some final particles. So
phases of amplitudes\ can be measured, [1,2].

Here I derive general formulations capable of fully describe the phenomena and
apply them to results of [1].

\textbf{Expression A} --- Amplitudes, being complex valued, can always be
expressed as
\begin{equation}
A=|A|e^{i\text{ }\phi}\ \ \ \text{and}\ \ \ \bar{A}=|\bar{A}|e^{i\text{ }%
\bar{\phi}}. \label{P0.0}%
\end{equation}
\noindent\ Direct CP violations are usually quantified by $\Delta_{cp}\neq0
$,
\begin{equation}
\Delta_{cp}\equiv(|A|^{2}-|\bar{A}|^{2})/(|A|^{2}+|\bar{A}|^{2})\text{,}
\label{dlt-R-CP}%
\end{equation}
\noindent where $A$ and $\bar{A}$ represent $A_{\left(  tot\right)  }$ and
$\bar{A}_{\left(  tot\right)  }$ or $A_{\left(  k\right)  }$ and $\bar
{A}_{\left(  k\right)  }$ in Eqs.(\ref{Asum}). (The symbol $\Delta_{cp}$ is
used, in stead of $\mathcal{A}_{cp},$ to avoid confusion with amplitudes.)

$\Delta_{cp}$ is insensitive to the phase of $\bar{A}/A$, which is convention
dependent. To describe phase CP violation we should use amplitudes, denoted by
$\bar{A}^{\prime}$ and $A^{\prime}$, which have the phase convention such that
CP invariant amplitudes satisfy $\bar{A}_{inv}^{\prime}/A_{inv}^{\prime
}=1e^{i0}$. Then
\begin{equation}
\bar{A}^{\prime}/A^{\prime}=R_{cp}\text{ }e^{i\Phi_{cp}}\equiv Z_{cp},\text{
\ \ }-\pi<\Phi_{cp}\leq\pi, \label{Zcp}%
\end{equation}
and their deviations from\ $Z_{cp,inv}=1e^{i0}$ give full quantifications of
direct CP violations, modulus and phase.

\textbf{Expression B} --- Belle [1] used another model-independent expressions
for $B^{-}$ and $B^{+}$ respectively,%
\begin{align}
A^{\prime}  &  =ae^{i\delta_{B}}\text{(}1-be^{i\varphi}\text{), \ }\bar
{A}^{\prime}=ae^{i\delta_{B}}\text{(}1+be^{i\varphi}\text{),}\label{B}\\
\Delta_{cp}  &  =-2b\cos\varphi/\left(  1+b^{2}\right)  . \label{dlta-B}%
\end{align}
I denote $B_{cp}\equiv be^{i\varphi}$and obtain MT conformal relations
\begin{align}
\bar{A}^{\prime}/A^{\prime}  &  \equiv Z_{cp}=(1+B_{cp})/(1-B_{cp}%
),\label{Zcp-B}\\
B_{cp}  &  =-(1-Z_{cp})/(1+Z_{cp}). \label{B-Zcp}%
\end{align}
\textbf{\ } Of the one-to-one and onto properties (circles/lines
$\Leftrightarrow$ circles/lines) of MT, I point out some highlights.
$b=0\Leftrightarrow Z_{cp}=1e^{i0}$. So $b\neq0$ gives CP violation, in
modulus and phase allocated by $b$ and $\varphi$. [$0<b<1$, $\varphi=_{0}%
^{\pi}$]$\Leftrightarrow\Phi_{cp}=$ $0$, thus all in $\Delta_{cp}%
=\pm2b/\left(  1+b^{2}\right)  $; [$b\neq0$, $\varphi=\pm\pi/2$]
$\Leftrightarrow\Delta_{cp}=0$ ($R_{cp}=1$), thus all in $\Phi_{cp}%
=\pm2\arctan b$; maximal $\Phi_{cp}=$ $\pi$ at [$1<b$, $\varphi=_{0}^{\pi}$];
maximal $\Delta_{cp}=\pm1$ at [$b=1,$ $\varphi=_{0}^{\pi}]$, where
[$R_{cp}=_{\infty}^{0}$, $\Phi_{cp}$ arbitrary].

Belle [1] assumed the nonresonant parts to be CP invariant and measured all
$be^{i\varphi}$ and $\delta_{B}$. I calculate $Z_{cp}=R_{cp}$ $e^{i\Phi_{cp}}$
(versus only $\Delta_{cp}=\mathcal{A}_{cp}$ calculated in [1]), thus revealing
their measurements of direct CP violations both in the modulus and in the
phase, shown at the end with other quantities I derive after giving the
realization of Expression B in the KM framework [7].

\textbf{Direct CP violation in the KM framework\textbf{\ ---}} Direct CP
violations come about naturally in the KM framework, as first established
theoretically in $K$ mesons (the $s$ particles) [8]. Many other references and
discussions can be found in the reviews of Particle Data Group (PDG) [12-14].
Here I give a self-contained discussion.

Weak decay amplitudes\ without particle-antiparticle oscillation are expressed
as
\begin{align}
A  &  =V_{ub}V_{us}^{\ast}A_{1}+V_{tb}V_{ts}^{\ast}A_{2}\text{ }\equiv
z_{1}\text{\ }A_{1}+z_{2}\text{\ }A_{2},\label{KM-A}\\
\bar{A}  &  =V_{ub}^{\ast}V_{us}A_{1}+V_{tb}^{\ast}V_{ts}A_{2}\equiv
z_{1}^{\ast}\text{\ }A_{1}+z_{2}^{\ast}\text{\ }A_{2}\text{ ,} \label{KM-Abar}%
\end{align}
first for the $b\rightarrow s$ decays and then for decays with $z_{1}$ and
$z_{2}$ as elements from the CKM matrix $\mathbb{V}$, [4,7].

One of the attributes of $\mathbb{V}$ is unitarity:%
\begin{align}%
{\textstyle\sum\nolimits_{m=1}^{3}}
V_{u_{m}d_{j}}V_{u_{m}d_{k}}^{\ast}  &  =\delta_{jk}\text{, \ \ }%
j,k=1,2,3\text{,}\label{dd'}\\%
{\textstyle\sum\nolimits_{j=1}^{3}}
V_{u_{m}d_{j}}V_{u_{n}d_{j}}^{\ast}  &  =\delta_{mn}\text{, \ \ }%
m,n=1,2,3\text{,} \label{tt'}%
\end{align}
where the letters $\left(  u_{m},d_{j}\right)  $ denote the weak isospin
doublets: ($u,d$), ($c,s$), ($t,b$) quarks [or ($\nu_{e},e$), ($\nu_{\mu},\mu
$), ($\nu_{\tau},\tau$) leptons involving the PMNS matrix, [15].]

Eqs.(\ref{KM-A},\ref{KM-Abar}) can be derived by drawing quark diagrams and
combing terms with the same $z$. Because of $\sum\nolimits_{l=1}^{3}z_{l}=0$
conditions in Eqs.(\ref{dd'},\ref{tt'}), $A$ and $\bar{A}$ can always be
expressed by two terms as in Eqs.(\ref{KM-A},\ref{KM-Abar}) and in different
ways. The strong amplitudes $A_{1},A_{2}$ contain strong interactions to all
orders. The relative phase of particle and antiparticle states is chosen such
that the same strong amplitudes $A_{1},A_{2}$ appear in $A$, $\bar{A}.$ I will
show that a suitable choice $\mathbb{V}^{\prime}$ can be made to obtain
$A^{\prime}$, $\bar{A}^{\prime}$ (related to $A$, $\bar{A}$ by a phase
transformation) so that phase direct CP violations can be quantified. That
will be Expression C.

In [16], I did a comprehensive study of direct CP violation for $c$, $b$, and
$s$ particle decays. Writing
\begin{equation}
\Delta_{cp}=\frac{-4\operatorname{Im}\left(  z_{1}^{\ast}z_{2}\right)
\operatorname{Im}\left(  A_{1}^{\ast}A_{2}\right)  }{|z_{1}\text{\ }%
A_{1}+z_{2}\text{\ }A_{2}|^{2}+|z_{1}^{\ast}\text{\ }A_{1}+z_{2}^{\ast
}\text{\ }A_{2}|^{2}}, \label{dlt-KM}%
\end{equation}
I found that all $\operatorname{Im}\left(  z_{1}^{\ast}z_{2}\right)  =$ $\pm
c_{1}c_{2}c_{3}\left(  s_{1}\right)  ^{2}s_{2}s_{3}s_{\delta},$ in the
notation of [7]. (This was four years before [17], whose parametrization has
been called by PDG [12] the standard parametrization for the CKM and the PMNS
matrices, thanks to the "advocation" and use by [14, 15].) The unique and
ubiquitous $|\operatorname{Im}\left(  z_{1}^{\ast}z_{2}\right)  |$ found in
$\Delta_{cp}$ were denoted by the symbol%
\begin{equation}
X_{cp}\equiv|\operatorname{Im}\left(  z_{1}^{\ast}z_{2}\right)  |=c_{12}%
\left(  c_{13}\right)  ^{2}c_{23}s_{12}s_{13}s_{23}s_{\alpha_{13}},
\label{Xcp-CK}%
\end{equation}
in [17,18]. I have been using it since. It touches upon aspects and
developments of the theory complimentary to those the symbol $J$ does,
[12-15]. It serves as a reminder that its relevance to experiments is through
its role in direct CP violations.

Dividing the numerator and the denominator in Eq.(\ref{dlt-KM}) by
$|z_{1}||z_{2}||A_{1}||A_{2}|$ (assuming none of them are zero for now) and
simplifying, we obtain%
\begin{equation}
\Delta_{cp}=-2\sin\theta\sin\Theta/\text{(}l+l^{-1}+2\cos\theta\cos
\Theta\text{),} \label{dlt-KM-C}%
\end{equation}
where
\begin{align}
\sin\theta &  \equiv\operatorname{Im}(z_{1}^{\ast}z_{2})/|z_{1}z_{2}|,\text{
\ \ }r=|z_{2}|/|z_{1}|,\text{ }\nonumber\\
\sin\Theta &  =\operatorname{Im}\left(  A_{1}^{\ast}A_{2}\right)  /|A_{1}%
A_{2}|,\text{ \ }R=|A_{2}|/|A_{1}|;\text{ or}\nonumber\\
re^{i\theta}  &  =z_{2}/z_{1}\equiv z_{21},\text{ \ }R\text{ }e^{i\Theta
}=A_{2}/A_{1}\equiv A_{21}; \label{KM-defs}%
\end{align}
and $l\equiv rR.$ The various $|\sin\theta|$ are\
\begin{align}
\sin\alpha_{d_{j}d_{k}}  &  =X_{cp}/|\text{(}V_{ud_{j}}V_{ud_{k}}^{\ast
}\text{)(}V_{td_{j}}V_{td_{k}}^{\ast}\text{)}|,\label{sinA-dd'}\\
\sin\beta_{d_{j}d_{k}}  &  =X_{cp}/|\text{(}V_{td_{j}}V_{td_{k}}^{\ast
}\text{)(}V_{cd_{j}}V_{cd_{k}}^{\ast}\text{)}|,\label{sinB-dd'}\\
\sin\gamma_{d_{j}d_{k}}  &  =X_{cp}/|\text{(}V_{cd_{j}}V_{cd_{k}}^{\ast
}\text{)(}V_{ud_{j}}V_{ud_{k}}^{\ast}\text{)}|, \label{sinG-dd'}%
\end{align}
and similarly for $\sin\alpha_{u_{m}u_{n}}$, $\sin\beta_{u_{m}u_{n}}$, and
$\sin\gamma_{u_{m}u_{n}}$. (In the case of $d_{j}d_{k}$ being $bd$ , the
$\alpha,\beta,\gamma$ notations conform to those in [12].) Each set of
$\alpha,\beta,\gamma$ with the subscripts $d_{j}d_{k}$ (or $u_{m}u_{n}$) is
associated with the $d_{j}d_{k}$ (or $u_{m}u_{n}$) orthogonal relation of
Eqs.(\ref{dd'},\ref{tt'}), thus the $d_{j}d_{k}$ (or $u_{m}u_{n}$) triangle on
the complex plane. To get the signs of various $\sin\theta,$ it is best to use
a specific parametrization, like the standard parametrization or its
variations (which are needed for reasons to be discussed). Amplitudes of a
particular set of decays, Eqs.(\ref{KM-A},\ref{KM-Abar}), involve one
particular triangle; yet, once CP violation is established in one decay (as
has been) all $|z|\neq0$ and all $\sin\theta\neq0$.

\textbf{Variations to the standard parametrization in the standard
construction }--- To define $Z_{cp}$ and realize Expression B in the KM
framework, I first show that the $z_{1}$ for a chosen $A_{1}$ can be made real
and positive by using a suitable parametrization. [Note that all $A_{\left(
k\right)  }$and $\bar{A}_{\left(  k\right)  }$ in Eq.(\ref{Asum}) can be made
to have the same $z_{1}$].

In [17], besides the standard parametrization of $\mathbb{V}$, Keung and I
found (by trials) a construction procedure for it: $\mathbb{V}=\mathbb{R}%
\left(  23\right)  \mathbb{U}\left(  13\right)  \mathbb{R}\left(  12\right)
$, one factor for each independent plane. $\mathbb{R}\left(  jk\right)  $ is
the rotation matrix in the $jk$-plane and $\mathbb{U}\left(  jk\right)  $ is
$\mathbb{R}\left(  jk\right)  $ with $\pm s_{jk}\rightarrow\pm s_{jk}e^{\mp
i\alpha_{jk}}$,
\begin{equation}
\text{ }\mathbb{U}\left(  13\right)  \equiv\left(
\begin{array}
[c]{ccc}%
c_{13} & 0 & s_{13}e^{-i\alpha_{13}}\\
0 & 1 & 0\\
-s_{13}e^{i\alpha_{13}} & 0 & c_{13}%
\end{array}
\right)  . \label{dlta13-V}%
\end{equation}

\noindent(Symbols $\alpha_{jk}$ are used, saving $\delta_{jk}$ for the
Kronecker deltas.) So the standard parametrization is $R\left(  23\right)
U\left(  13\right)  R\left(  12\right)  $-parametrization. The procedure also
provides variations: $\mathbb{V}^{\prime}=\mathbb{U}\left(  23\right)
\mathbb{R}\left(  13\right)  \mathbb{R}\left(  12\right)  ,$ or $\mathbb{V}%
^{\prime\prime}=\mathbb{R}\left(  23\right)  \mathbb{R}\left(  13\right)
\mathbb{U}\left(  12\right)  $, or $\mathbb{V}s$ with different ordering of
$\left(  23\right)  \left(  13\right)  \left(  12\right)  $. For $b\rightarrow
s$ , $b\rightarrow d$, and $s\rightarrow d$ decays, $\mathbb{V}^{\prime}$
gives real positive $z_{1}^{\prime}=$ $V_{ub}^{\prime}V_{us}^{\prime\ast}$,
$V_{ub}^{\prime}V_{ud}^{\prime\ast}$, or $V_{ud}^{\prime}V_{us}^{\prime\ast}$.

Here I digress to give a fuller explanation of the above and formulate (what I
would "advocate" to call) the \textit{standard construction}s for $N\times N$
CKM, PMNS, and general unitary matrices. Let us start with the following.

\textbf{The Murnaghan construction of }$N\times N$\textbf{\ general}
\textbf{unitary matrices }[19]:%
\begin{equation}
\mathbb{U}=\mathbb{FA}\text{, \ \ \ where \ \ \ }\mathbb{A\equiv}%
{\textstyle\prod\limits_{j<k\leq N}}
\mathbb{U}\left(  jk\right)  , \label{Mnn}%
\end{equation}
$\mathbb{F}$= diag($e^{i\phi_{1}},e^{i\phi_{2}},\cdots,e^{i\phi_{N}}$) and the
$\frac{1}{2}N\left(  N-1\right)  $ number (one for each plane in $N$
dimensions) of $\mathbb{U}\left(  jk\right)  $ are defined above
Eq.(\ref{dlta13-V}). Different orderings of $\mathbb{U}\left(  jk\right)  $
give different (equally valid) parametrizations of $\mathbb{U}.$

$\mathbb{U}$ given by Eqs.(\ref{Mnn}) has all the attributes of a $N\times N$
unitary matrix. For example, there are $\frac{1}{2}N\left(  N-1\right)  $
angles and [$\frac{1}{2}N\left(  N-1\right)  +N$]= $\frac{1}{2}N\left(
N+1\right)  $ phases.

\textbf{Theorem 1:} The core matrix $\mathbb{C}$, obtained from
$\mathbb{A}$ by the maximal-phase-stripping similarity
transformation (ST):
\begin{equation}
\mathbb{C}=\mathbb{DAD}^{\dagger}=%
{\textstyle\prod\limits_{j<k\leq\left(  N-1\right)  ,\text{ }m<N}}
\mathbb{U}^{\prime}\left(  jk\right)  \mathbb{R}\left(  mN\right)
\label{C-mtx}%
\end{equation}
with
\begin{equation}
\mathbb{D}=diag(e^{i\alpha_{1N}},e^{i\alpha_{2N}},\cdots,e^{i\alpha_{\left(
N-1\right)  N}},1), \label{Phs-1Tr}%
\end{equation}
has the least possible number of phases under ST:
[$\frac{1}{2}N\left( N-1\right)  -\left(  N-1\right)
$]$=\frac{1}{2}\left(  N-1\right)  \left( N-2\right)  $.

To prove the theorem, stick $\mathbb{D}^{\dagger}\mathbb{D}$
in-between all $\mathbb{U}\left(  jk\right)  $ of $\mathbb{A}$ in
Eq.(\ref{Mnn}) and see that $\mathbb{DU}\left(  jk\right)
\mathbb{D}^{\dagger}=\mathbb{U}^{\prime}\left( jk\right)  $ with
$\alpha_{jk}^{\prime}=\alpha_{jk}-\alpha_{jN}+\alpha_{kN}$, giving
$\alpha_{jN}^{\prime}=0$ and $\mathbb{U}^{\prime}\left(  jN\right)
=\mathbb{R}\left(  jN\right)  $. So the ST of Eq.(\ref{C-mtx}) maxes
out the phase stripping from
$\mathbb{A}$ [and all $\mathbb{A}$ with different ordering of $\mathbb{U}%
\left(  jk\right)  $]. \textbf{Corollary 1}: The phases in a core
matrix $\mathbb{C}$ can be moved around by phase-moving ST (see an
example of it later).

Using $\mathbb{C}$\ we can make the following explicit constructions. Let us
call them the standard constructions.

\textbf{The standard construction of }$N\times N$\textbf{\ general}
\textbf{unitary matrices }[revealing more phase structures than the Murnaghan
construction, Eq.(\ref{Mnn})]:%
\begin{equation}
\mathbb{U}=\mathbb{F\mathbb{D}^{\dagger}CD\equiv F}^{\prime}\mathbb{CD}%
\text{.} \label{Ustd}%
\end{equation}

\textbf{The standard construction of }$N\times N$\textbf{\ CKM matrices for
quarks}:\textbf{\ }%
\begin{equation}
\mathbb{V=C}^{q}. \label{KM-Udirac}%
\end{equation}
Theorem 1 and \ Eq.(\ref{Ustd}) show by explicit construction how the usual
phase counting works out. ($2N-1$) out of the $2N$ phase\ freedoms of quark
fields are used to strip away all that can be from the up-down quark mixing
matrix $\mathbb{U}^{q}$ by phase transformations (PT): $\mathbb{C}%
^{q}\mathbb{=F}^{\prime\dagger}\mathbb{U}^{q}\mathbb{D}^{\dagger}$,
Eq.(\ref{Ustd}) to Eq.(\ref{KM-Udirac}). So, always one phase is left free. It
can be used to make one (only one) of the many $A_{1}$ in Eq.(\ref{Asum})
real. In K decays, setting zero-isospin-change amplitude real is the Wu-Yang
phase convention [6].

What Keung and I found by trials in [17] is the $3\times3$
forerunner of this standard construction and [18] extended it to a
$4\times4$ case. An example of Corollary 1 is the phase-moving ST,
$\mathbb{V}^{\prime}$=
diag($1,1,e^{-i\alpha_{13}}$)$\mathbb{V}$diag($1,1,e^{i\alpha_{13}}$)
and
$\alpha_{23}^{\prime}$=$-\alpha_{13}$. Other such relations among $\mathbb{V}%
$, $\mathbb{V}^{\prime}$, $\mathbb{V}^{\prime\prime}$ are left as exercises.

\textbf{The standard construction of }$N\times N$\textbf{\ PMNS matrices}
\textbf{for} \textbf{Dirac leptons and for} \textbf{Majorana }$\nu$:
\begin{equation}
\mathbb{V}^{l_{D}}=\mathbb{C}^{l_{D}},\ \ \ \ \text{and}\ \ \ \ \mathbb{V}%
^{\nu_{M}}=\mathbb{C^{\nu_{M}}D}. \label{PMNS}%
\end{equation}

\noindent Dirac lepton fields have the same phase freedoms as quarks fields,
so $\mathbb{V}^{l_{D}}$\ is given by a core matrix as is $\mathbb{V}$ for
quarks. However, Majorana neutrino fields do not have phase freedom [15], so
only the phase freedoms of the Dirac leptons can be used to strip away $N$
phases from the Dirac-Majorana mixing matrix $\mathbb{U^{\nu_{M}}}$ by one PT:
$\mathbb{C^{\nu_{M}}D=F}^{\prime\dagger}\mathbb{U^{\nu_{M}}}$, Eq.(\ref{Ustd})
to Eq.(\ref{PMNS}). What has been adapted in neutrino research [15] is the
$3\times3$ case of [17] for $\mathbb{C^{\nu_{M}}}$. All variations discussed
here can also apply.

In Eq.(\ref{Mnn}) I can also use $\mathbb{U}=\mathbb{A}^{\prime}\mathbb{F}$
with $\mathbb{A}^{\prime}\equiv\mathbb{FAF}^{\dagger}$, follow similar
procedure and obtain another core matrix $\mathbb{C}^{\prime}=\mathbb{D}%
^{\prime}\mathbb{A}^{\prime}\mathbb{D}^{\prime\dagger}$ for $\mathbb{U}$. I
also have theorems that give different constructions with core matrices
involving less than $\frac{1}{2}N\left(  N-1\right)  $ planes, like the Euler
construction for $SO\left(  3\right)  $. However, I see no advantage over the
standard construction for the uses discussed here. Further, I can use the core
matrices to give spectral constructions for matrices. I give details of these
results in [20].

\textbf{Expression C }--- Representing CP invariant amplitudes\ by $A_{1}%
\neq0$ and using the $\mathbb{V}^{\prime}$ in which $z_{1}^{\prime}=|z_{1}|$,
I obtain Expression C:%
\begin{align}
A^{\prime}  &  =|z_{1}|A_{1}(1+z_{21}A_{21})=de^{i\delta_{1}}\text{(}%
1+le^{i\left(  \Theta+\theta\right)  }\text{)},\text{ }\label{C-A}\\
\bar{A}^{\prime}  &  =|z_{1}|A_{1}(1+z_{21}^{\ast}A_{21})=de^{i\delta_{1}%
}\text{(}1+le^{i\left(  \Theta-\theta\right)  }\text{),} \label{C-Abar}%
\end{align}
where $de^{i\delta_{1}}\equiv|z_{1}|A_{1}$ and relations given by
Eqs.(\ref{KM-defs}) still hold -- good exercise to check; and
\begin{align}
\bar{A}^{\prime}/A^{\prime}  &  \equiv Z_{cp}=(1+z_{21}^{\ast}A_{21}%
)/(1+z_{21}A_{21}),\label{Zcp-C}\\
A_{21}  &  =(1-Z_{cp})/(z_{21}-z_{21}^{\ast}Z_{cp}). \label{A21-C}%
\end{align}
Now the CP invariant $\bar{A}_{inv}^{\prime}/A_{inv}^{\prime}=1e^{i0}$. What
we have done above is equivalent (and gives justification) to starting with
$\mathbb{V}$ and making the phase change to amplitudes:\ $A^{\prime
}=e^{-i\text{ }\alpha_{1}}A$ and $\bar{A}^{\prime}=e^{i\text{ }\alpha_{1}}%
\bar{A}$ with $e^{i\text{ }\alpha_{1}}\equiv z_{1}/|z_{1}|$.

$Z_{cp}$ is in terms of the knowables (measurable in principle), $z_{21}$ and
$A_{21}$, and is TM conformally related to $A_{21}$ (not $z_{21}$). Their
one-to-one and onto mapping properties carry important information.
Im$z_{21}|A_{21}|=l\sin\theta=0\Leftrightarrow Z_{cp}=1e^{i0}$. So
$l\sin\theta\neq0$ assures CP violation, in modulus or phase allocated by
$l\sin\theta$ and\ $\Theta$. $\Delta_{cp}$ is still given by
Eq.(\ref{dlt-KM-C}). Since all $|z|\neq0$ and all $\sin\theta\neq0$ (i.e.,
Im$z_{21}=r\sin\theta\neq0$ and well defined), so in all decays direct CP
violations happen everywhere on the whole $A_{21}$ complex plane except one
point, $A_{21}=0$. [For example: at $\Theta=0$, $\Delta_{cp}=0$ and $\Phi
_{cp}=-2\arctan$[$l\sin\theta/$($1+l\cos\theta$)]; and at $\Theta=\pi\pm
\theta$ and $l=1$, $\Delta_{cp}=\pm1$.] Further, for any value of
Im$z_{21}=r\sin\theta\neq0$, CP violation can be large if $A_{21}$ cooperates.
Eq.(\ref{A21-C}) gives the $A_{21}$ for a $Z_{cp}$ asked for.

Besides giving the conceptual understanding mentioned above and the
realization of Expression B to be discussed below, Expression C also gives the
possibility of finding $z_{21},$ $A_{21}$ and $|z_{1}|A_{1}$ from data via
Eqs.(\ref{C-A},\ref{C-Abar}); and provides versatile ways of analyzing data.
[If the data are not sensitive to these many parameters, one can put in
$z_{21}$ from [12] and find $A_{21}$ and $|z_{1}|A_{1}$.]

\textbf{Realization of Expression B in the KM framework and derivation of
amplitudes from Belle [1] } --- Using Eqs.(\ref{C-A},\ref{C-Abar}) of
Expression C, decomposing $e^{\pm i\theta}$ into real and imaginary parts,
identifying%
\begin{align}
ae^{i\delta_{B}}  &  =|z_{1}|A_{1}\text{(}1+A_{21}r\cos\theta\text{),
\ }\label{KM-a-term}\\
B_{cp}  &  =-iA_{21}r\sin\theta/\left(  1+A_{21}r\cos\theta\right)  ,\text{\ }
\label{KM-b-termp}%
\end{align}

\noindent I obtain the realization of Expression B, Eqs.(\ref{B}), in terms of
knowables in the KM framework.

Besides $\bar{A}^{\prime}/A^{\prime}\equiv Z_{cp}=R_{cp}$ $e^{i\Phi_{cp}}$
mentioned earlier, Eqs.(\ref{Zcp-B}), I now can derive from
Eqs.(\ref{KM-a-term}, \ref{KM-b-termp})%
\begin{align}
A_{21}  &  =-B_{cp}/[ir\sin\theta+B_{cp}r\cos\theta]\text{ \ \ and}%
\label{strong21}\\
A_{1}  &  =ae^{i\delta_{B}}/[|z_{1}|\text{(}1+A_{21}r\cos\theta\text{)}]
\label{strong1}%
\end{align}
in terms of $B_{cp}$. The MT conformal relations between $B_{cp}$ and $A_{21}
$, Eqs.(\ref{KM-b-termp},\ref{strong21}),\ are anticipated from the MT
relations that $Z_{cp}$ has both with $B_{cp}$, Eqs.(\ref{Zcp-B},\ref{B-Zcp}),
and with $A_{21}$, Eqs.(\ref{Zcp-C},\ref{A21-C}).

Substituting into Eqs.(\ref{KM-a-term},\ref{KM-b-termp}) $\delta_{B}$ and
$B_{cp}\equiv be^{i\varphi}$ from [1], and $\sin\theta=-\sin\alpha_{bs}%
\approx-0.82,$ $\cos\theta\approx\allowbreak-0.57,$ and $r=|V_{tb}V_{ts}%
^{\ast}|/|V_{ub}V_{us}^{\ast}|\approx46$ (derived using [12,14] and assuming
all angles in the standard parametrization of $\mathbb{V}$ being in the first
quadrant), I obtain the values of $A_{21}\equiv A_{2}/A_{1}$ and $\tilde
{A}_{1}\equiv(|z_{1}|/a)A_{1}$, listed below together with $Z_{cp}=R_{cp}$
$e^{i\Phi_{cp}}$ for four of the decays observed by [1] as examples: $B^{\mp}$
to \{1\} $K^{\ast}\left(  892\right)  \pi^{\mp},$ \{2\} $K^{\ast}\left(
1430\right)  \pi^{\mp}$, \{3\} $\rho^{0}\left(  770\right)  K^{\mp}$, \{4\}
$f_{2}\left(  1270\right)  K^{\mp}$,
\begin{align*}
\text{\{1\} }Z_{cp}  &  =1.16\exp\text{(-}i0.048\text{)},\ \ \\
A_{21}  &  =0.0021\exp\text{(-}i1.8\text{)},\text{ }\tilde{A}_{1}%
=0.98\exp\text{(-}i0.052\text{);}\\
\text{\{2\} }Z_{cp}  &  =0.93\exp\text{(-}i0.12\text{)},\ \ \\
\ A_{21}  &  =0.0019\exp\text{(}i2.\,\allowbreak5\text{)},\text{\ }\tilde
{A}_{1}=0.96\exp\text{(}i0.99\text{);}\\
\text{\{3\} }Z_{cp}  &  =0.74\exp\text{(-}i0.46\text{)},\ \ \ \\
A_{21}  &  =0.0087\exp\text{(}i2.\,\allowbreak4\text{)},\text{ }\tilde{A}%
_{1}=0.85\exp\text{(-}i0.24\text{);}\\
\text{\{4\} }Z_{cp}  &  =1.98\exp\text{(-}i0.34\text{)},\ \\
A_{21}  &  =0.011\exp\text{(-}i1.7\text{)},\text{ \ }\tilde{A}_{1}%
=0.93\exp\text{(}i2.2\text{).}%
\end{align*}
Only central values are shown. The proper way to find errors\ in $R_{cp}$ and
$\Phi_{cp}$ is to analyze data distributions in $Z_{cp}$ by authors of Belle
[1].\ However, I did carry out various error calculations using statistical
errors in $b$ and $\varphi$ given by [1] and noticed the following. When an
error in $\varphi$ decreases (increases) modulus CP violation, it increases
(decreases) phase CP violation; in contrast, when an error decreases
(increases) $b$, both the modulus and the phase CP violations decrease
(increase). The phase CP violation, $\Phi_{cp}\neq0$, in case \{2\} stood out, [21].

$A_{1}=(a/|z_{1}|)\tilde{A}_{1}$ can be derived once Belle publishes values of
$a$, using partial rates and $f_{k}$ of Eq.(\ref{Asum}). (Note the wide range
of the central values of the moduli and phases of $\tilde{A}_{1}$ and $A_{21}%
$. The proper way to obtain them\ and their error analyses will be to fit data
using Expression C, [21].) These $A_{1}$ and $A_{2}$ from experiments can be
compared with theory. (For current theoretical calculation schemes, see
[22,23], e.g.). Alternatively, use $A_{1}$ and $A_{2}$\ from theory in
Eqs.(\ref{KM-a-term},\ref{KM-b-termp}), then solve for $r$ and $\theta,$ and
compare them with those obtained elsewhere.

\textbf{Conclusion} --- The formulations given here have general applications
for studying phase and modulus direct CP violations and strong amplitudes in
weak decays, beyond the results calculated here for $B^{\mp}\rightarrow
K^{\mp}\pi^{\pm}\pi^{\mp}$ of [1]. The M\"{o}bius (linear fractional
conformal) transformation relations found here tell us that in the KM
formulation, once a CP violation is established in one reaction (as has been),
the amount of it (phase and modulus) in other decays is unrestricted by the
CKM matrix, but solely dependent on how cooperative the strong amplitudes are.
This\ new understanding is encouraging for the search of direct CP violations
in general. The versatile procedure given here for the constructions of
$N\times N$ CKM, PMNS, and general unitary matrices clarifies the $3\times3$
cases and is useful for the beyond.

\end{document}